\documentclass{pazhb}
\usepackage[hyperfootnotes=false]{hyperref}
\usepackage{color}
\usepackage{epsf}
\usepackage[utf8]{inputenc}
\usepackage[russian]{babel}
\usepackage[T2A]{fontenc}
\usepackage{amsmath}
\usepackage{amsfonts}
\usepackage{amssymb}
\usepackage{epsf}
\usepackage{graphicx}
\usepackage{gensymb}
\usepackage{float}
\usepackage{longtable}
\usepackage{changes}
\usepackage{wrapfig}
\begin{document}

\newcommand{\obja}{J0854}
\newcommand{\objb}{J1344}

\journalinfo{2023}{0}{0}{1}[0]
\UDK{///}

\title{SDSS J085414.02+390537.3 --- a new asynchronous polar}
\author{
A.I.~Kolbin\email{kolbinalexander@mail.ru}\address{1,2,3}, 
M.V.~Suslikov\address{1,2}, 
V.Yu.~Kochkina\address{1,2}, 
N.V.~Borisov\address{1}, 
A.N.~Burenkov\address{1}, 
D.V.~Oparin\address{1}
\addresstext{1}{Special Astrophysical Observatory of RAS, Nizhnij Arkhyz, Karachay-Cherkessian rep., 369167, Russia}
\addresstext{2}{Kazan (Volga-region) Federal University, Kremlevskaya str. 18, Kazan 420008, Russia}
\addresstext{3}{North Caucasian Federal University, Pushkina str. 1, Stavropol 355017, Russia}
}


\shortauthor{A.I.~Kolbin, et al.}

\shorttitle{SDSS J085414.02+390537.3 --- a new asynchronous polar}

\begin{abstract}
The asynchrony of the polar SDSS~J085414.02+390537.3 was revealed using the data of ZTF photometric survey. The light curves show a beat period $P_{beat} = 24.6 \pm 0.1$~days. During this period the system changes its brightness by $\sim 3^m$. The periodograms show power peaks at the white dwarf's rotation period $P_{spin} = 113.197 \pm 0.001$~min and orbital period $P_{orb} = 113.560 \pm 0.001$~min with the corresponding polar asynchrony $1-P_{orb}/P_{spin} = 0.3\%$. The photometric behavior of the polar indicates a change of the main accreting pole during the beat period. Based on the Zeeman splitting of the H$\beta$ line, an estimate of the mean magnetic field strength of the white dwarf is found to be $B = 28.5\pm 1.5$~MG. The magnetic field strength $B \approx 34$~MG near the magnetic pole was found by modeling cyclotron spectra. Doppler tomograms in the H$\beta$ line demonstrate the distribution of emission sources typical for polars.

\keywords{Stars: novae, cataclysmic variables -- Individual: SDSS~J085414.02+390537.3 -- Methods: photometry, spectroscopy.}
\end{abstract}

\section{INTRODUCTION}

Cataclysmic variables are close binary systems with orbital periods $P_{orb} = 1.4-9$~h, which consist of a white dwarf and a low-mass cold star (usually an M-dwarf) filling its Roche lobe \citep{Warner95, Hellier01}. The matter of the cold component flows out from the inner Lagrange point L$_1$ and, in the case of a weak magnetic field of the white dwarf ($B\lesssim 0.1$~MG), forms an accretion disk.
Another picture of accretion is observed in systems with a strong magnetic field ($B\gtrsim 10$~MG), where ionized gas flows onto the surface of the accretor along magnetic lines without the formation of an accretion disk. When the incident gas interacts with the surface of the accretor, hot ($T\sim 10$~keV) and compact accretion spots are formed. They are sources of hard X-ray radiation, as well as polarized cyclotron radiation in the optical and near infrared ranges. Objects of this kind are called AM~Her stars or polars \citep{Cropper90}. In polars, the interaction of the strong magnetic field of the accretor with the donor leads to synchronization of the rotation of the white dwarf with its orbital orbital motion ($P_{spin} = P_{orb}$, where $P_{spin}$ is the period of the white dwarf's rotation). At magnetic fields $B\sim 0.1-10$~MG, an accretion disk is formed, which is destroyed from the inside by the white dwarf's magnetic field. These systems are called DQ~Her stars or intermediate polars \citep{Patterson94}. Unlike polars, systems of the DQ~Her type are not synchronous, and the rotation period of a white dwarf is on average $0.1P_{orb}$.

Among the polars, a small group of objects is distinguished, called asynchronous polars. These systems show a weak asynchrony not exceeding a few percent. To date, asynchrony has been confirmed for V1500~Cyg \citep{Pavlenko18}, V1432~Aql \citep{Littlefield15}, BY~Cam \citep{Silber92}, CD~Ind \citep{Littlefield19}, SDSS J084617.11+245344.1 \citep{Littlefield23}, 1RXS J083842.1-282723 \citep{Halpern17}, IGR~J19552+0044 \citep{Tovmassian17}. It is assumed that this state is unstable and asynchronous systems move towards a state of synchronous motion. It is possible that such systems were brought out of synchronous motion by the recent explosion of Nova \citep{Stockman88}. According to modern concepts, the distinguishing feature of asynchronous polars is their temporary presence in the state of asynchrony, while for intermediate polars the state with $P_{spin}<P_{orb}$ should be stable \citep{King91}. The difference between the orbital period and the rotation period of the white dwarf leads to a periodic change in the geometry of accretion flows, the drift of accretion spots on the surface of the accretor, and a change in the main accreting magnetic pole \citep{Sobolev21}. This leads to manifestation in asynchronous polars of periodicity at the beat frequency $\omega - \Omega$, where $\omega = 1/P_{spin}$ is the white dwarf's rotation frequency, and $\Omega = 1/P_{orb}$ is the orbital frequency.

The object SDSS J085414.02+390537.3 (hereafter {\obja}) was suspected by \cite{Christian01} to be a  magnetic cataclysmic variable. \cite{Szkody05} discovered cyclotron harmonics in the {\obja} spectrum, their position corresponded to a magnetic field strength of 44~MG. A high circular polarization of the optical radiation, reaching 30~\%, was also observed. \cite{Dillon08} determined the photometric period of the polar $P = 113.26 \pm 0.03$~min.  The light curve had a two-peak bright phase, which was interpreted as the passage of an accretion spot across the disk of the white dwarf. We became interested in {\obja} after discovering signs of asynchrony in the long-term light curves of the ZTF survey. For a more detailed study of this system, we analyzed its phase resolved spectroscopy.

This work is structured as follows. In the second section, we describe the performed spectral observations {\obja} and their reduction. Further, in the third section, we analyze the long-term light curves {\obja} obtained by the ZTF survey. There we study the variability on the beat period, as well as the variability modulated by the rotation of the white dwarf. In the fourth section, the {\obja} spectra are analyzed, including the determination of the magnetic field from the Zeeman splitting of the H$\beta$ line, modeling of the cyclotron spectrum, and Doppler tomography. The results of the work are summarized in the section ``Conclusion''.

\section{2. Observations and data reduction}

The sets of {\obja} spectra were obtained at the night of March 07–08, 2022 and April 24–25, 2022 with the 6-m BTA telescope of the Special Astrophysical Observatory of the Russian Academy of Sciences. Observations were carried out using the SCORPIO--2 and SCORPIO--1 \citep{Afan} focal reducers for the first and second nights, respectively. The longslit spectroscopy was performed with exposures of 300~s. During the March observations, only $\approx 0.6$ of the polar's orbital period was covered, but in April, a total period of $\approx 110$~min was observed. In the March observations, the spectra covered the range 3700--7300~\AA\, with a resolution $\Delta \lambda \approx 5.2$~\AA. The observations were made under light cloudiness and the seeng of $2.5''$. At the April observations, the spectra werte obtained in 3800--5700~\AA\, range with a resolution $\Delta \lambda \approx 5$~\AA. These observations were made under good astroclimatic conditions with a seeng of $1.8''$.


The observation data was reduced using the IRAF\footnote{The IRAF astronomical data processing and analysis package is available at https://iraf-community.github.io.} software package. The bias frame was subtracted from the spectral images and they were flat-fielded to reduce a microvariation of CCD sensitivity. Cosmic ray traces were removed using the LaCosmic \citep{Dokkum} algorithm. The correction for geometric distortions and wavelength calibration were carried out using the frames of the He-Ne-Ar lamp. The spectra were optimally extracted \citep{Horne86} with sky background subtraction. The spectrophotometric calibration was performed by observation of the Feige~34 (for March 07/08, 2022) and AGK+81$^{\circ}$266 (for April 24/25, 2022) standards. The fluxes were corrected for the variability of atmospheric opacity from the spectra of the neighboring star captured by the spectrograph slit. Barycentric Julian Dates (BJD) and barycentric corrections for radial velocity were calculated for each spectrum. 

\begin{figure*}
  \centering
	\includegraphics[width=0.8\textwidth]{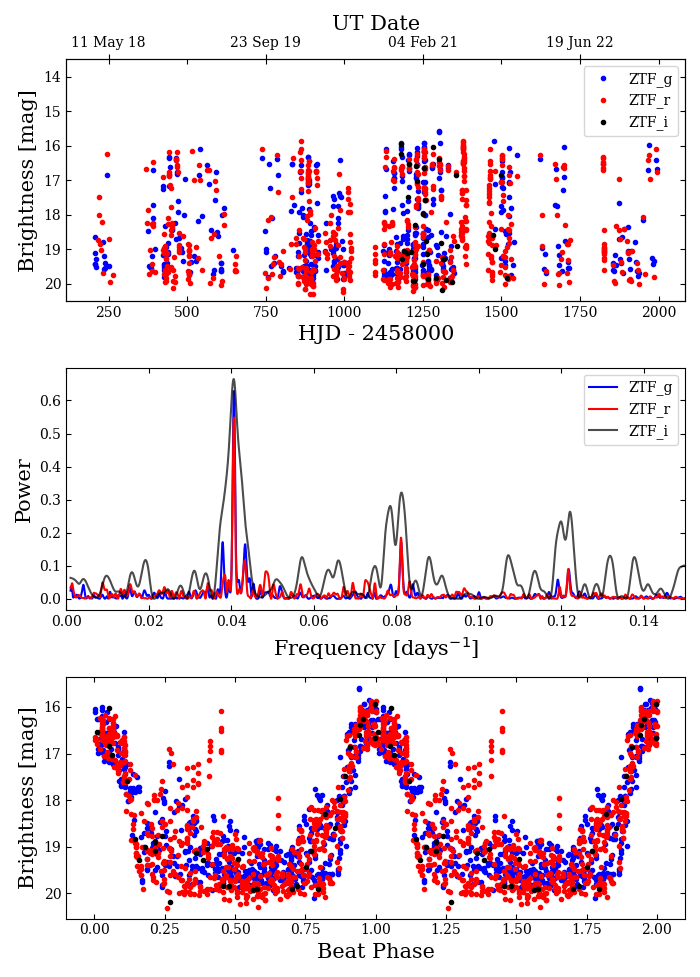}
\caption{Upper panel: long-term {\obja} light curves obtained by the ZTF survey in the $g$, $r$, $i$ bands. Middle panel: Lomb--Scargle periodograms obtained for light curves in three bands. Bottom panel: phased light curves plotted from ephemeris (\ref{ephem_long}). 
}
\label{fig:phot_long}
\end{figure*} 

\section{3. Analysis of ZTF photometry}

\subsection{Variability at beat frequency}

The long-term light curves of {\obja} obtained by ZTF \citep{masci18} survey over almost 4.8 years in the $g$, $r$, $i$ bands are shown in Fig. \ref{fig:phot_long}. One can see rather fast (on the order of a month) changes in the brightness of the object from $20^m$ to $16^m$ in three filters. The same figure shows the Lomb--Scargle \citep{VanderPlas18} periodograms derived from the presented light curves. Noteworthy is the power peak at the frequency $f=0.04060 \pm 0.00023$~day$^{-1}$ (period $P=24.63\pm0.14$~day), which appears from the observational data in three filters. On Fig. \ref{fig:phot_long} phase light curves foldered for the found period are also presented. They show a high state with an average brightness of $\langle g \rangle \approx \langle r \rangle \approx \langle i \rangle \approx 16.5^m$ and a low state with $\langle g \rangle \approx \langle r \rangle \approx \langle i \rangle \approx 19.5^m$. The periodicity of the brightness state is beyond doubt, but it is atypical for systems of the AM~Her type. Obviously, the obtained period significantly exceeds the orbital periods of the polars ($\sim 1/10$~days), and the observed brightness variability in $\approx 3^m$ noticeable exceeds the amplitude of the out--of--eclipse orbital variability of the polars ($\sim 1^m$). The amplitude $\sim 3^m$ is typical for the change in the states of the polars due to the variability of the accretion rate. However, such switching of brightness states is irregular and does not show any significant periodicity. We assumed that {\obja} is an asynchronous polar, and the found period is the beat period $P_{beat} = (\omega-\Omega)^{-1}$. In other words, this period is equal to the rotation period of the white dwarf in the rotating coordinate system of the binary. The observed brightness variability in this case can be caused by a change in the main accreting magnetic pole (see the ``Conclusion'' section for more details). Note that the phased light curves were constructed according to ephemeris
\begin{equation}
    HJD_{beat} = 2459082.49(7) + 24.6(1) \times E,
\label{ephem_long}
\end{equation}
where the zero epoch corresponds to the middle of the high state.

\begin{figure*}
  \centering
	\includegraphics[width=\textwidth]{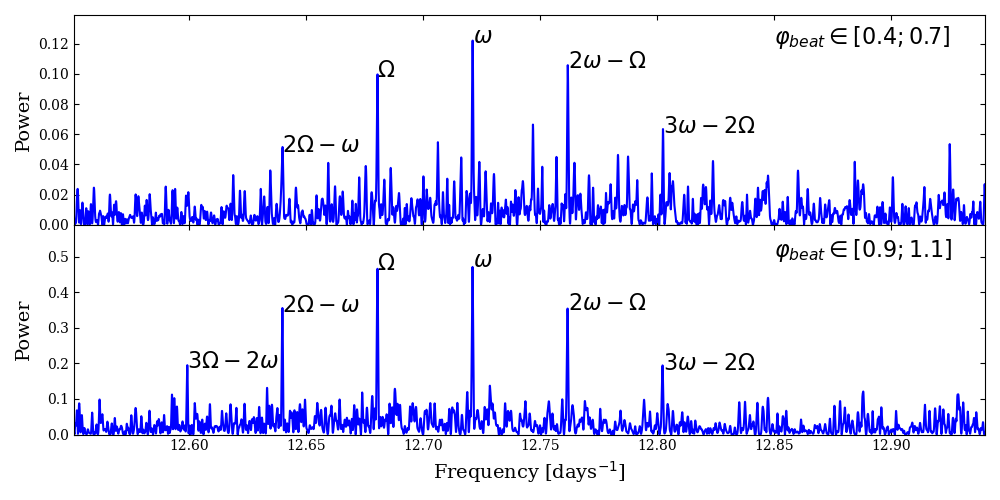}
\caption{The periodograms of {\obja} near the white dwarf rotation frequency for low ($\varphi_{beat} = 0.4-0.7$, upper panel) and high ($\varphi_{beat} = 0.9-1.1$, lower panel) states. The positions of the peaks of the white dwarf's rotation frequency $\omega$, the orbital frequency of $\Omega$, and the sidebands are indicated.
}
\label{fig:periodogramm}
\end{figure*} 

\subsection{Orbital and spin variability}

To analyse the fast variability, two phase ranges $\varphi_{beat}=0.4-0.7$ and $\varphi_{beat}=0.9-1.1$ were selected, corresponding to the low and high states, respectively (see Fig. \ref{fig:phot_long}). These regions have small light dispersions, which gives us hope to identify in them a rotational or orbital variability undistorted by the accretion rate change. The analysis was carried out in the bands $g$, $r$ with preliminary subtraction of the mean luminosity. In the region $\varphi_{beat}=0.4-0.7$ the mean brightness was considered constant, and in the region $\varphi_{beat}=0.9-1.1$ it strongly depends on $\varphi_{beat}$ and was found by approximation of the light curve section with a parabola. The Lomb-Scargle periodograms for the two sites are shown in Fig. \ref{fig:periodogramm}. They contain many peaks, the strongest of which corresponds to the period $P_{spin}=113.197\pm 0.001$~min, close to the rotation period of the white dwarf \cite{Dillon08}. The frequency of the neighbouring peak is $\omega-1/P_{beat}$ and is the orbital frequency of $\Omega$, and its corresponding orbital period is $P_{orb} = 0.078861\pm 0.000001$~day ($113.55984 \pm 0.001$~min). Other power peaks are sidebands, produced by combining the $\omega$ and $\Omega$ frequencies.

\begin{figure}[h!]
  \centering
	\includegraphics[width=\linewidth]{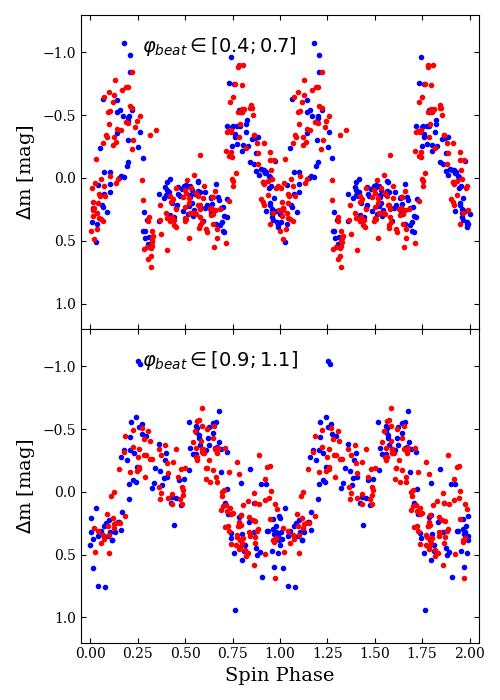}
\caption{Phased light curves of {\obja} derived from ephemeris (\ref{ephem_dip}) and showing the variability modulated with the rotation of the white dwarf. The upper panel shows light curves derived from low state data $\varphi_{beat} = 0.4-0.7$, and the lower panel --- from high state data $\varphi_{beat} = 0.9-1.1$. The $g$-band measurements are shown by blue dots, and the $r$-band measurements are shown by red dots. 
}
\label{fig:phot_short}
\end{figure} 

The light curves in the low ($\varphi_{beat} = 0.4-0.7$) and high ($\varphi_{beat} = 0.9-1.1$) states foldered with the found rotation period of the white dwarf are shown in Fig. \ref{fig:phot_short}. In both bands, they have a double-peaked bright phase that appears to be formed during the accretion spot's passage across the white dwarf's disc. The characteristic double-peak structure of the bright phase is due to the beaming of the cyclotron emission from the accretion spot. It is most intense when the spot is near the edge of the white dwarf disc, i.e., when the angle between the magnetic field lines and the line of sight is close to $90^{\circ}$ (if the rotation axis of the star has a high inclination, see, e.g., \cite{Kolbin20, Kolbin22}). The ephemeris
\begin{equation}
HJD_{rot} = 2459063.376(3) + 0.0786091(7)\times E
\label{ephem_dip}
\end{equation}
were used to plot the presented phased light curves. The zero epoch corresponds to the middle of the bright phase for $\varphi_{beat} = 0.4-0.7$.

We note the change in the position of the bright phase at $\approx {^1/_2} P_{spin}$ during the transition from the low to the high state. This phenomenon fits well into the picture of {\obja} asynchrony and is apparently caused by the change of the main accreting magnetic pole.

\section{4. Spectrum analysis}

\subsection{Zeeman splitting}

\begin{figure*}
\centering
\includegraphics[width=\textwidth]{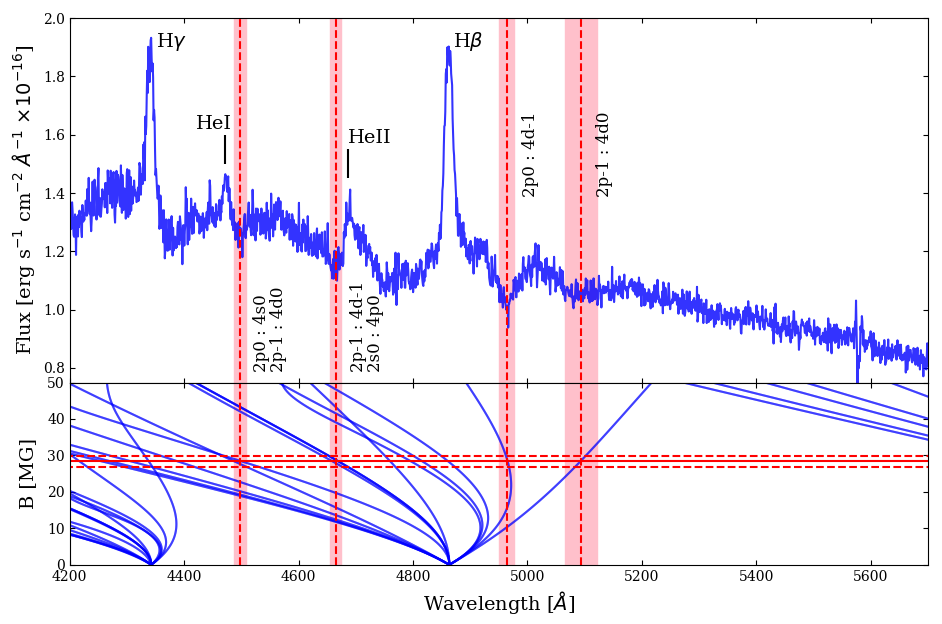}
\caption{Top panel: averaged spectrum of {\obja} obtained from observations on April 24/25, 2022. The vertical lines show the positions of the Zeeman splitting components of the H$\beta$ line, and the pink bars indicate their uncertainty. The identified transitions in the hydrogen atom are marked. Bottom panel: the diagram of Zeeman splitting for H$\alpha$, H$\beta$, H$\gamma$ lines. The solid horizontal line corresponds to the magnetic field estimate.}
\label{fig:zeeman}
\end{figure*}

The {\obja} spectra of April 24/25, 2022 contain the absorption components of the Zeeman splitting of the H$\beta$ line.
The Zeeman splitting is typical for polars at a low accretion rates, when the white dwarf emission reveals in the spectra. The low accretion rate at the time of observations of {\obja} confirms the absence of an intense HeII~$\lambda$4686 line. Fig. \ref{fig:zeeman} shows the averaged {\obja} spectrum with well-identified Zeeman components. The central wavelength of the Zeeman components was determined by a Gaussian approximation and the value of $1\sigma$ was used as an error of component position. To estimate the magnetic field of a white dwarf, we calculated the wavelengths of transitions in the hydrogen atom in a strong magnetic field using the program code \cite{Schimeczek14}. The resulting diagram of Zeeman splitting for the H$\alpha$, H$\beta$, H$\gamma$ lines is also shown in Fig. \ref{fig:zeeman}. The average magnetic field of the white dwarf is estimated as $B=28.5\pm 1.5$~MG.

\subsection{Cyclotron spectrum}

\begin{figure*}
\centering
\includegraphics[width=\textwidth]{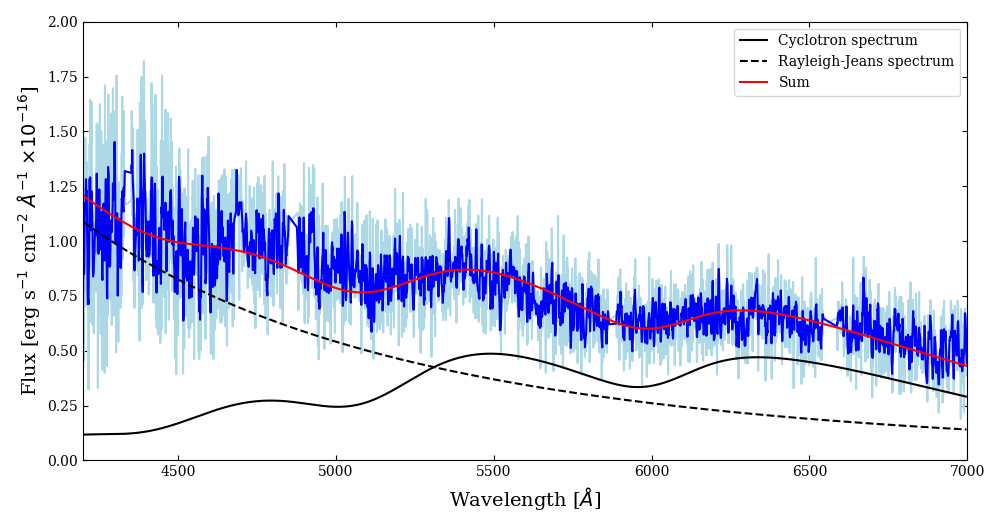}
\caption{The {\obja} spectrum with cyclotron harmonics (blue curve, emission lines removed) and its approximation (red curve) by the sum of the cyclotron spectrum (black curve) and the Rayleigh-Jeans (dashed black curve). The blue curve shows the observed spectrum smoothed by the Savitzky–-Golay filter.}
\label{fig:cyclotron}
\end{figure*}

From the spectral set obtained on March 07/08, 2022, the spectrum with weak cyclotron harmonics is noteworthy (see Fig. \ref{fig:cyclotron}). We analysed it with a simple model if an accretion spot which is uniform in temperature and density. This model is often used in the study of polars \citep{Campbell08, Kolbin19, Beuermann20}. The Rayleigh-Jeans spectrum was added to the cyclotron spectrum to take into account the white dwarf emission and the possible contribution of the accretion flow. The model depends on four parameters: the magnetic field strength in the spot $B$, the electron temperature $T_e$, the angle between the magnetic field lines with respect to the line of sight $\theta$, and the plasma parameter $\Lambda = \omega^2_p \ell / \omega_c c$, where $\omega_p$ is the plasma frequency, $\ell$ is the depth of the emitting region along the line of sight, and $\omega_c = eB/m_e c$ is the cyclotron frequency. The computation of the intensity of cyclotron radiation was performed using the absorption coefficients calculated by the \cite{Chan81} method. The best fit of the observed spectrum corresponds to the magnetic field in the spot $B = 34$~MG and the temperature $T_e=10$~keV. This value of the magnetic field is $\approx 6$~MG greater than the estimate from the Zeeman splitting. Such discrepancies between the two methods are typical for AM~Her type systems. When analyzing the photospheric lines of a white dwarf, the average (over the observed surface of the star) magnetic strength is determined, while the slightly higher magnetic strength near the magnetic pole is determined from cyclotron harmonics. 

\begin{figure*}
\centering
\includegraphics[width=\textwidth]{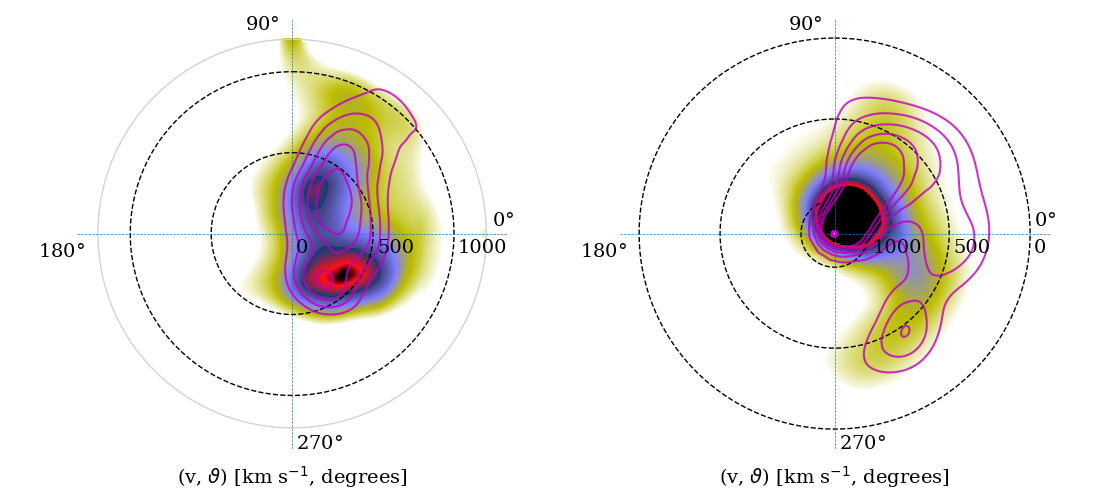}
\caption{Doppler tomograms of {\obja} in H$\beta$ line. On the left is the tomogram in the standard projection, while on the right is the tomogram in the inside--out projection. The tomograms reconstructed from the spectra on March 07/08, 2022 are presented in color scale, and the tomograms on April 24/25, 2022 are displayed by isolines.}
\label{fig:dopptom}
\end{figure*}

\subsection{Doppler tomography}

Doppler tomograms represent the distribution of emission line sources in a two-dimensional velocity space. Each point of this space is determined by two polar coordinates: the absolute velocity relative to the center of mass of the system $v$ (up to the factor $\sin i$, where $i$ --- is the orbital inclination) and the angle $\vartheta$, which determines the direction of the velocity vector in the orbit plane (usually this angle is measured from the line connecting the centers of mass of the stellar components). For details of  Doppler tomograms interpretation, we refer the reader to \cite{Marsh16, Kotze15, Kotze16}. Since we do not have {\obja} orbital ephemeris, the tomograms are reconstructed up to rotation angle, i.e. the angle $\vartheta$ of tomogram points is defined up to a constant. The {\obja} tomography was performed for the H$\beta$ line which has the maximum signal-to-noise ratio in both sets of observations. Doppler tomograms were reconstructed using the \cite{Kotze15} program code, which implements the maximum entropy method.

The resulting {\obja} tomograms are shown in Figs. \ref{fig:dopptom} in two projections: standard and inside--out. In the first case, the velocity $v$ increases from the center of the tomogram to its outer edge. In the inside--out projection, on the contrary, the absolute velocity increases from the edge to the center. The last option is preferable for studying high-speed structures that would be smeared over a large area in the standard projection (see for details \cite{Kotze15}). For the convenience of comparing tomograms obtained on different nights, we presented the intensity distribution according to the data of April 24/25, 2022 as a set of isolines. We superimposed these isolines on the map obtained from the spectra on March 07/08, 2022, presented in color scale.

The {\obja} tomograms are typical for AM~Her type systems. There is an intensity maximum in the fourth quadrant ($270^{\circ} \le \vartheta < 360^{\circ}$) and a weaker bright area in the first quadrant ($0^{\circ} \le \vartheta < 90^{\circ}$), which is elongated in the radial direction. The same picture appears on the map in the inside--out projection, but differs in a more pronounced high-velocity component. As we noted the Doppler maps are reconstructed up to rotation angle, and we could rotate them at $\sim 180^{\circ}$ and see correspondence with maps of other polars, for example, HU~Aqr \citep{Schwope97}, BS~Tri \citep{Kolbin22}.

The phases of the beat period at the times of spectral observations are $\varphi_{beat} = 0.91\pm 0.09$ and $\varphi_{beat} = 0.87\pm 0.1$ for the March and April observations, respectively. The high phase uncertainty is due to the beat period error. The two series of spectral observations differ only by $\Delta \varphi_{beat} = 0.04$. It seems that such a phase difference is not sufficient to highlight differences in the accretion flow. Although Doppler tomograms also have different positions of intensity maxima (this is especially noticeable on maps in the standard projection), these differences can be caused by incomplete coverage of the orbital period in March observations ($\approx 0.6P_{orb}$).

\section{Conclusion}

\begin{figure}[h]
  \centering
	\includegraphics[width=\linewidth]{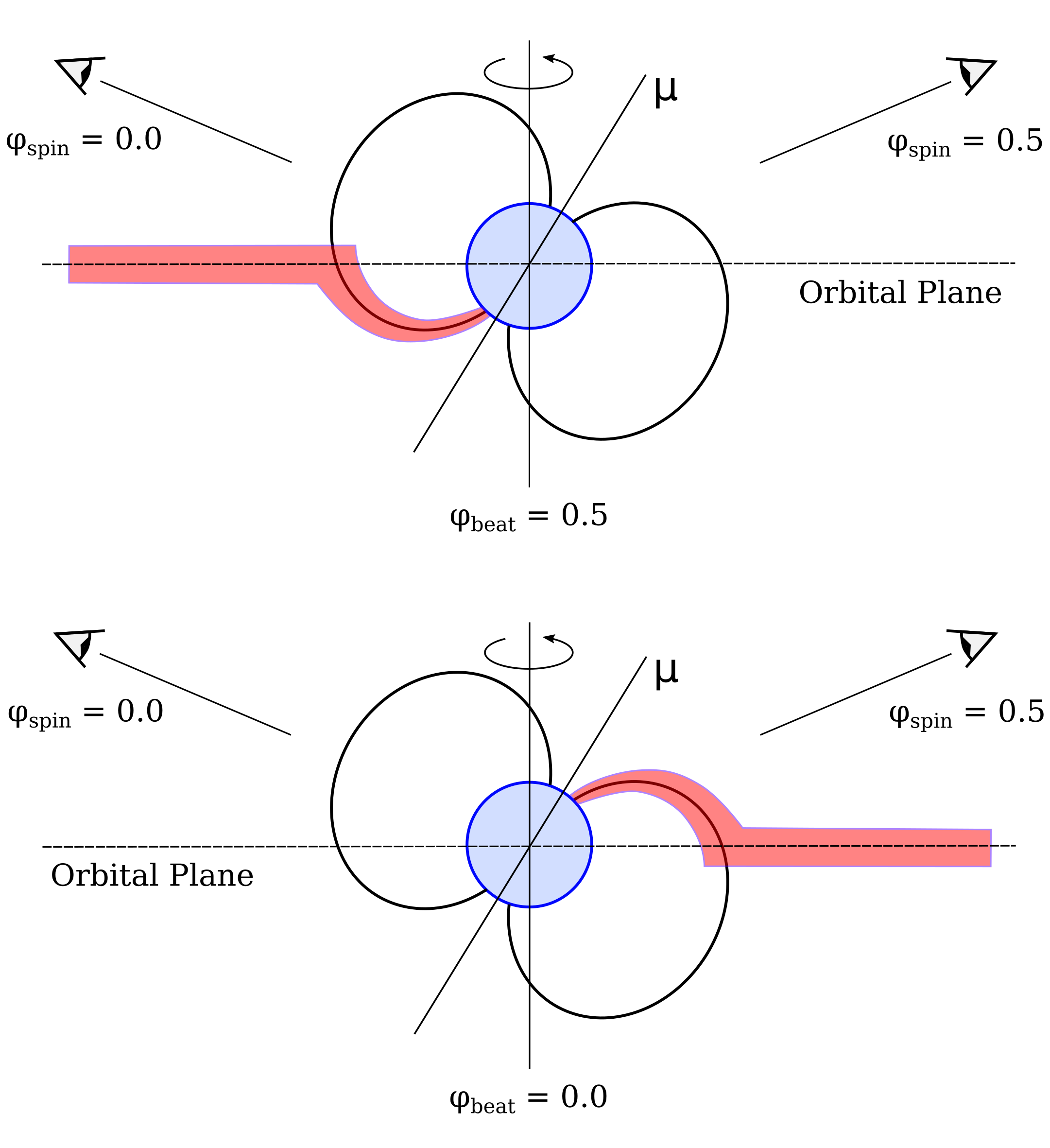}
\caption{On the interpretation of photometric observations of {\obja}. Accretion onto a white dwarf is shown in low ($\varphi_{beat} = 0.5$) and high ($\varphi_{beat} = 0.0$) states. In the low state, accretion occurs near the pole of rotation, turned away from the observer. After half the beat period, the magnetic dipole is oriented towards the incoming gas with the opposite magnetic pole, which has better visibility conditions. In this case, the rotation phase of the white dwarf, on which the accreting pole is visible, changes from $\varphi_{spin} = 0$ to $\varphi_{spin} = 0.5$.}
\label{fig:scheme}
\end{figure}

In the present work, a new asynchronous polar {\obja} has been discovered. In its long-term light curves obtained by the ZTF survey, a beat period $P_{beat}=24.6\pm0.1$~day is distinguished. During this period, the polar changes its mean brightness from $\langle g \rangle \approx 19.5^m$ (low state) to $\langle g \rangle \approx 16.5^m$ (high state). An analysis of the light curves in the low and high states revealed the rotational variability of the white dwarf with a period $P_{spin} = 113.197 \pm 0.001$~min. The periodograms of the low and high states also show an orbital period $P_{orb} = 113.55984 \pm 0.001$~min and sidebands arising from the modulation of rotational variability by the orbital motion. The asynchrony of {\obja} is equal to $(\Omega-\omega)/\Omega \approx 0.3\%$ and is consistent with the values of this parameter for asynchronous polars, where it is $<2\%$. The {\obja} light curve foldered with a rotation period in the low state has a two-hump bright phase, which is interpreted by the passage of a source of cyclotron radiation (i.e., an accretion spot) across the disk of a white dwarf. A similar shape of the light curve is also observed in the high state, but it is shifted by about half the rotation period. This effect can be interpreted as a change in the main accreting pole when the state changes from low to high. Since the average brightness of the polar in the high state is higher by $\sim 3^m$, it can be assumed that the accreting pole in it has better visibility conditions, i.e. located near the pole of rotation facing the observer. In addition, the differences in the brightness of the two states can be related to the possible difference in the magnetic field strength at the magnetic poles. In this case, since in the high state the brightness does not fall to the level of the low state, it is likely that the accretion spot does not go beyond the disk of the white dwarf. The latter means that the inclination of the dipole to the axis of rotation is $\beta<90^{\circ}-i$, where $i$ is the inclination of the axis of rotation to the line of sight. The described scheme of accretion in {\obja} is shown in Fig. \ref{fig:scheme}.

Zeeman components of the H$\beta$ line splitting are present in the {\obja} spectra. They were used to obtain an estimate of the magnetic field of the white dwarf $B=28.5\pm 1.5$~MG. Cyclotron harmonics were detected in one of the spectra, the position of which corresponds to a magnetic field $B \approx 34$~MG. The higher estimate of the magnetic field strength is due to the fact that it is obtained for the region close to the magnetic pole, while the Zeeman splitting is used to find the magnetic field strength averaged over the surface of the star. Doppler tomograms show a distribution of emission line sources typical of polars, showing low-velocity structures likely to form near the ballistic trajectory, as well as high-speed features likely originating in the white dwarf's magnetosphere. Significant differences in tomograms obtained in different periods of time were not found. Apparently, this is due to the small difference in the beat phase between the spectral series.

The polar {\obja} is an interesting object for further optical observations. To test the hypothesis of switching between the main accreting poles, polarization observations in different phases of the beat period are desirable. When accretion changes from one pole to another, one should expect a change in the sign of circular polarization. It would be interesting to carry out additional phase-resolved spectral observations. The corresponding Doppler maps would make it possible to study changes in the geometry of accretion flows during the rotation of the magnetic dipole relative to the donor.

{\bf Acknowledgments.} The study was supported by the Russian Science Foundation grant No. 22-72-10064, https://rscf.ru/project/22-72-10064/. Observations with the SAO RAS telescopes are supported by the Ministry of Science and Higher Education of the Russian Federation. The renovation of telescope equipment is currently provided within the national project ''Science and universities.''


\begin{thebibliography}{16}


\bibitem[\protect\citeauthoryear{Afanasiev \& Moiseev}{2011}]{Afan}
V.L. Afanasiev, A.V. Moiseev.
{\journal{BaltA}{20}{363}{2011}}

\bibitem[\protect\citeauthoryear{Beuermann, Burwitz, et al.}{2020}]{Beuermann20} 
K. Beuermann, V. Burwitz, K. Reinsch, et al.
{\journal{\aap}{634}{91}{2020}}

\bibitem[\protect\citeauthoryear{Campbell, Harrison, et al.}{2008}]{Campbell08} 
R.K. Campbell, T.E. Harrison, A.D. Schwope, et al.
{\journal{\apj}{672}{531}{2008}}

\bibitem[\protect\citeauthoryear{Chanmugam \& Dulk}{1981}]{Chan81}
G. Chanmugam, G.A. Dulk.
{\journal{\apj}{244}{569}{1981}}

\bibitem[\protect\citeauthoryear{Christian, Craig, et al.}{2001}]{Christian01}
D.J.~Christian, N.~Craig, J.~Dupuis, et al.
{\journal{IBVS}{5032}{1}{2001}}

\bibitem[\protect\citeauthoryear{Cropper}{1990}]{Cropper90}
M. Cropper.
{\journal{\ssr}{54}{195}{1990}}

\bibitem[\protect\citeauthoryear{Dillon, B.~T. G{\"a}nsicke, et al.}{2008}]{Dillon08}
M. Dillon, B.~T. G{\"a}nsicke, et al.
{\journal{\mnras}{386}{1568}{2008}}

\bibitem[\protect\citeauthoryear{van Dokkum}{2001}]{Dokkum}
P.G. van Dokkum.
{\journal{\pasp}{113}{1420}{2001}}

\bibitem[\protect\citeauthoryear{Halpern, Bogdanov, et al.}{2017}]{Halpern17}
J.P.~Halpern, S.~Bogdanov, J.R.~Thorstensen.
{\journal{\apj}{838}{124}{2017}}

\bibitem[\protect\citeauthoryear{Hellier}{2001}]{Hellier01}
С. Hellier.
{Cataclysmic Variable Stars (Springer)}

\bibitem[\protect\citeauthoryear{Horne}{1986}]{Horne86}
K. Horne.
{\journal{\pasp}{98}{609}{1986}}

\bibitem[\protect\citeauthoryear{King \& Lasota}{1991}]{King91}
A.R.~King, J.P.~Lasota
{\journal{\apj}{378}{674}{1991}}

\bibitem[\protect\citeauthoryear{Kolbin, Serebryakova, et al.}{2019}]{Kolbin19}
A.I. Kolbin, N.A. Serebryakova, M.M. Gabdeev, et al.
{\journal{\apbul}{74}{80}{2019}}

\bibitem[\protect\citeauthoryear{Kolbin \& Borisov}{2020}]{Kolbin20}
A.I. Kolbin, N.V. Borisov.
{\journal{\astl}{46}{812}{2020}}

\bibitem[\protect\citeauthoryear{Kolbin, Borisov, et al.}{2022}]{Kolbin22}
A.I. Kolbin, N.V. Borisov, N.A. Serebriakova, et al.
{\journal{\mnras}{511}{20}{2022}}

\bibitem[\protect\citeauthoryear{Kotze, Potter, et al.}{2015}]{Kotze15}
E.J. Kotze, S.B. Potter, V.A. McBride
{\journal{\aap}{579}{77}{2015}}

\bibitem[\protect\citeauthoryear{Kotze, Potter, et al.}{2016}]{Kotze16}
E.J. Kotze, S.B. Potter, V.A. McBride.
{\journal{\aap}{595}{47}{2016}}

\bibitem[\protect\citeauthoryear{Littlefield, Mukai, et al.}{2015}]{Littlefield15}
C.~Littlefield, K.~Mukai, R.~Mumme, et al.
{\journal{\mnras}{449}{3107}{2015}}

\bibitem[\protect\citeauthoryear{Littlefield, Garnavich, et al.}{2015}]{Littlefield19}
C.~Littlefield, P.~Garnavich, et al.
{\journal{\apj}{881}{141}{2019}}

\bibitem[\protect\citeauthoryear{Littlefield, Hoard, et al.}{2023}]{Littlefield23}
C.~Littlefield, D.W.~Hoard, P.~Garnavich, et al.
{\journal{\aj}{165}{43}{2023}}

\bibitem[\protect\citeauthoryear{Marsh \& Schwope}{2016}]{Marsh16}
T.R. Marsh, A.D. Schwope.
{\journal{ASSL}{439}{195}{2016}}

\bibitem[\protect\citeauthoryear{Masci, Laher, et al.}{2018}]{masci18}
F.~Masci, R.~Laher, B.~Rusholme, et al.
{\journal{\pasp}{131}{995}{2018}}

\bibitem[\protect\citeauthoryear{Patterson}{1994}]{Patterson94}
J.~Patterson.
{\journal{\pasp}{106}{209}{1994}}

\bibitem[\protect\citeauthoryear{Pavlenko, Mason, et al.}{2018}]{Pavlenko18}
E.P.~Pavlenko, P.A.~Mason, A.A.~Sosnovskij, et al.
{\journal{\mnras}{479}{341}{2018}}

\bibitem[\protect\citeauthoryear{Schimeczek \& Wunner}{2014}]{Schimeczek14}
C.~Schimeczek, G.~Wunner.
{\journal{\apjs}{212}{26}{2014}}

\bibitem[\protect\citeauthoryear{Schwope, Mantel, et al.}{1997}]{Schwope97}
A.D. Schwope, K.H. Mantel, K. Horne.
{\journal{\aap}{319}{894}{1997}}

\bibitem[\protect\citeauthoryear{Silber, Bradt, et al.}{1992}]{Silber92}
A.~Silber, H.V.~Bradt, M.~Ishida, et al.
{\journal{\apj}{389}{704}{1992}}

\bibitem[\protect\citeauthoryear{Sobolev, Zhilkin, et al.}{2021}]{Sobolev21}
A.V.~Sobolev, A.G.~Zhilkin, D.V.~Bisikalo, et al.
{\journal{\arep}{65}{392}{2021}}

\bibitem[\protect\citeauthoryear{Stockman, Schmidt, et al.}{1988}]{Stockman88}
H.S.~Stockman, G.D.~Schmidt, D.Q.~Lamb.
{\journal{\apj}{332}{282}{1988}}

\bibitem[\protect\citeauthoryear{Szkody, Henden, et al.}{2005}]{Szkody05}
P.~Szkody, A.~Henden, O.J.~Fraser, et al.
{\journal{\aj}{129}{2386}{2005}}

\bibitem[\protect\citeauthoryear{Tovmassian, Gonz´alez-Buitrago, et al.}{2017}]{Tovmassian17}
[G.~Tovmassian, D.~Gonz´alez-Buitrago, J.~Thorstensen, et al.
{\journal{\aap}{608}{A36}{2017}}



\bibitem[\protect\citeauthoryear{VanderPlas}{2018}]{VanderPlas18}
J.T.~VanderPlas
{\journal{\apjs}{236}{16}{2018}}












\bibitem[\protect\citeauthoryear{Warner}{1995}]{Warner95}
B. Warner.
{Cataclysmic Variable Stars (Cambridge Univ. Press, Cambridge)}


\end{thebibliography}
\end{document}